\pdfoutput=1
\documentclass[a4paper,12pt]{article}

\usepackage[
    margin=0.8in
    ]{geometry}

\usepackage{amsmath,amssymb,amsthm,amsfonts,mathtools,mathrsfs,mathdots,latexsym}
\usepackage[italicdiff]{physics}
\mathtoolsset{showonlyrefs,showmanualtags}
\numberwithin{equation}{section}
\allowdisplaybreaks[4]

\usepackage[
    bibstyle=phys,
    sorting=none,
    biblabel=brackets,
    citestyle=numeric-comp,
    pageranges=false,
    eprint=true,
    url=true,
    ]{biblatex}
\DeclareSourcemap{
	\maps[datatype=bibtex]{
		\map[overwrite=true]{
			\step[fieldsource=shortjournal]
			\step[fieldset=journal, origfieldval]
		}
	}
}
\AtEveryBibitem{
    \iffieldequalstr{type}{NoiNSPIRE}{
        \DeclareFieldFormat{url}{[\href{#1}{Link}]}
        }{\DeclareFieldFormat{url}{[\href{#1}{iNSPIRE}]}}
    \ifentrytype{article}{%
        \iffieldundef{eprinttype}{}{%
            \iffieldundef{journal}{\clearfield{year}}{}
            }
        }{}
    \clearfield{note}
    \clearfield{type}
    \clearname{editor}
    \clearlist{location}
}
\bibliography{reference}

\usepackage{hyperref}
\hypersetup{
    colorlinks=true,
    linktocpage=true,
    linkcolor=blue,
    urlcolor=blue,
    citecolor=blue,
    filecolor=blue,
    bookmarksopen=true,
    }

\usepackage{authblk}

\title{Singular field redefinition between Witten's string field theory and Witten's theory deformed by Ellwood invariant}
\author{Yuji Ando\thanks{E-mail: \href{mailto:yuji.ando.b5@tohoku.ac.jp}{yuji.ando.b5@tohoku.ac.jp}}}
\affil{
Mathematical Science Center for Co-creative Society, Tohoku University,\\
Sendai, Miyagi 980-0845, Japan}
\date{}

\begin{document}
\maketitle
\renewcommand\thefootnote{\arabic{footnote}}
\setcounter{footnote}{0}
\begin{abstract}
    We construct a field redefinition between Witten's string field theory and its deformation by the Ellwood invariant. This field redefinition is singular and does not imply physical equivalence between them. However, it allows us to formally transfer classical solutions of Witten's theory to solutions of the deformed theory. Although the resulting solutions are also generically singular and require careful examination of their physical interpretation, we show that the tachyon vacuum solution can be consistently transferred from Witten's theory to the deformed theory.
\end{abstract}
\thispagestyle{empty}
\newpage
\setcounter{page}{1}
\tableofcontents
\section{Introduction}
In open string field theory, D-brane configurations are expected to be described as classical solutions of the equations of motion. Witten's bosonic open string field theory \cite{Witten:1985cc} has a remarkably simple action which consists only of a kinetic term and a cubic interaction term. As a result, a large class of classical solutions has been constructed, e.g. \cite{Takahashi:2002ez,Schnabl:2005gv,Okawa:2006vm,Schnabl:2007az,Kiermaier:2007ba,Erler:2009uj,Erler:2010zza,Murata:2011ex,Masuda:2012kt,Murata:2011ep,Erler:2014eqa,Ishibashi:2016xak,Miwa:2017oxy,Hata:2019dwu,Hata:2019ybw,Erler:2019fye} (for reviews see \cite{Fuchs:2008cc,Okawa:2012ica,Erler:2019vhl}). By contrast, no nontrivial classical solutions have been found in closed string field theory. For example, Zwiebach's bosonic closed string field theory action \cite{Zwiebach:1992ie} contains infinitely many interaction vertices, which makes the construction of classical solutions quite challenging.

In order to find classical solutions in closed string field theory, various approaches have been proposed so far. In particular, recent works attempt to obtain educated guesses for classical solutions of closed string field theory by deforming Witten's theory into a theory whose action contains infinitely many interaction vertices by adding stubs \cite{Schnabl:2023dbv,Schnabl:2024fdx,Stettinger:2024hkp}. While these approaches provide valuable insights into the structure of closed string field theory, this paper pursues a more direct approach.

We propose a field redefinition that allows classical solutions of Witten's theory to be transferred to another theory. If a field redefinition exists between Witten's theory and a closed string field theory, classical solutions in the latter could in principle be obtained from known solutions in the former. Of course, since Witten's theory is not physically equivalent to any closed string field theory, such a field redefinition is expected to have pathological properties. Indeed, it is known that a superstring field theory can be related to a free theory by a field redefinition, although this field redefinition is improper \cite{Erler:2015uba}.

As a first step, in this paper we investigate the relation between Witten's theory and Witten's theory deformed by the Ellwood invariant \cite{Zwiebach:1992bw}\footnote{In this paper, we refer to this theory as the deformed theory.}. The action of the deformed theory is obtained by adding the Ellwood invariant as an open-closed interaction term to Witten's action. Although this theory is not a closed string field theory, it is capable of incorporating on-shell closed strings. To investigate the relation between the two theories, we focus on their homotopy structure. By constructing an appropriate $A_{\infty}$ morphism, we find a field redefinition between them. This field redefinition is expressed in terms of infinitely many multilinear maps, and it remains unclear whether it is well-defined. However, when we express the field redefinition in terms of the star product rather than homotopy algebra notation, we observe that the field redefinition is singular.

Moreover, since the field redefinition is singular, the resulting transferred solutions are generically singular. Hence, we must verify whether they satisfy the equation of motion in the strong sense \cite{Hata:2011ke}. Nevertheless, we show that the transferred tachyon vacuum solution is regular and satisfies the equation of motion in the deformed theory. Additionally, we show that the cohomology of the shifted BRS operator is empty, as expected for a tachyon vacuum.

This paper is organized as follows. In section \ref{review}, we briefly review homotopy algebra and coalgebra, and summarize how field redefinitions can be formulated in this framework. In section \ref{fieldredef}, we construct multilinear maps relating Witten's theory and the deformed theory. In section \ref{star}, we study the transfer of classical solutions and show that we can obtain the tachyon vacuum solution. In section \ref{summary}, we present the summary. Appendix \ref{homotopyop} contains details on the homotopy operator.

\section{Field redefinition and cohomomorphism\label{review}}
In this section, we briefly review the homotopy structure of the theory and examine how the desired field redefinition can be formulated in the language of homotopy algebra. See \cite{Hohm:2017pnh,Jurco:2018sby,Erler:2019loq,Erbin:2020eyc} for more details. We follow the notation of these references \cite{Erler:2019loq,Erbin:2020eyc}.

First, let us consider Witten's theory \cite{Witten:1985cc}. For the open string field $\Psi$, its action $S_{0}[\Psi]$ is constructed from the BRS operator $m_{1}$, the string product $m_{2}$ defining the Witten vertex and symplectic inner product $\omega$ induced from the BPZ inner product.
\begin{equation}
    S_{0}[\Psi]=\frac{1}{2}\omega(\Psi,m_{1}(\Psi))+\frac{1}{3}\omega(\Psi,m_{2}(\Psi,\Psi))
\end{equation}
Here $\omega$ satisfies
\begin{equation}
    \omega(\Psi_{1},\Psi_{2})=-(-1)^{\abs{\Psi_{1}}\abs{\Psi_{2}}}\omega(\Psi_{2},\Psi_{1}),\label{cyclic}
\end{equation}
where $\abs{\Psi}$ is defined as the ghost number $\mathrm{gh}(\Psi)$ minus one. This action has cyclic $A_{\infty}$ structure and $m_{1},m_{2}$ satisfy the $A_{\infty}$ relations \cite{Stasheff:1963a,Stasheff:1963b}
\begin{align}
    0&=m_{1}\qty(m_{1}\qty(\Psi_{1}))\label{Ainf:1},\\
    0&=m_{1}\qty(m_{2}\qty(\Psi_{1},\Psi_{2}))+m_{2}\qty(m_{1}\qty(\Psi_{1}),\Psi_{2})+(-1)^\abs{\Psi_{1}}m_{2}\qty(\Psi_{1},m_{1}\qty(\Psi_{2})),\label{Ainf:2}\\
    0&=m_{2}\qty(m_{2}\qty(\Psi_{1},\Psi_{2}),\Psi_{3})+(-1)^\abs{\Psi_{1}}m_{2}\qty(\Psi_{1},m_{2}\qty(\Psi_{2},\Psi_{3})).\label{Ainf:3}
\end{align}
In addition, $m_{1}$ and $m_{2}$ are cyclic with respect to the symplectic form $\omega$.
\begin{align}
    0&=\omega(m_{1}(\Psi_{1}),\Psi_{2})+(-1)^\abs{\Psi_{1}}\omega(\Psi_{1},m_{1}(\Psi_{2}))\label{cycAinf:1}\\
    0&=\omega(m_{2}(\Psi_{1},\Psi_{2}),\Psi_{3})+(-1)^\abs{\Psi_{1}}\omega(\Psi_{1},m_{2}(\Psi_{2},\Psi_{3}))\label{cycAinf:2}
\end{align}

Alternatively, these conditions can be reformulated in the coalgebra language. Given a multilinear map $C_{n}$, the corresponding coderivation $\vb{C}_{n}$ is defined as
\begin{equation}
    \vb{C}_{n}\coloneqq\sum_{k=0}^\infty\qty(\sum_{l=0}^{k}\mathbb{I}^{\otimes k-l}\otimes C_{n}\otimes\mathbb{I}^{\otimes l})\pi_{n+k},
\end{equation}
where $\mathbb{I}$ is the identity map on the string state space $\mathcal{H}$. Following this definition, the multilinear maps $m_{1},m_{2}$ define coderivations $\vb{m}_{1},\vb{m}_{2}$ respectively and we write the sum of $\vb{m}_{1}$ and $\vb{m}_{2}$ as $\vb{m}$.
\begin{equation}
    \vb{m}\coloneqq\vb{m}_{1}+\vb{m}_{2}
\end{equation}
We introduce the bra notation $\bra{\omega}$ for the symplectic inner product, defined by $\omega(\Psi_{1},\Psi_{2})=\bra{\omega}\Psi_{1}\otimes\Psi_{2}$. With this notation, the $A_{\infty}$ relations together with the cyclicity conditions,
\eqref{cycAinf:1} and \eqref{cycAinf:2}, can be succinctly expressed as
\begin{equation}
    (\vb{m})^2=0\qc\bra{\omega}\pi_{2}\vb{m}=0.
\end{equation}
We further introduce a coderivation $\vb*{\partial}_{t}$ associated with the derivative along a path $\Psi(t)$. Using this notation, the action can be written as
\begin{equation}
    S_{0}[\Psi]=\int_{0}^{1}\dd{t}\bra{\omega}\pi_{1}\vb*{\partial}_{t}\frac{1}{1-\Psi(t)}\otimes\pi_{1}\vb{m}\frac{1}{1-\Psi(t)},
\end{equation}
where $\Psi(1)=\Psi,\Psi(0)=0$.

Similarly, the action of Witten's theory deformed by the Ellwood invariant can be reformulated in the language of homotopy algebra. Its action is given by adding Ellwood invariant to Witten's theory action. Here the Ellwood invariant is defined in terms of an on-shell closed string field $\hat{m}_{0}$ as follows.
\begin{equation}
    \omega(\Psi,\hat{m}_{0})
\end{equation}
Thus the action of deformed theory is given by
\begin{align}
    S_{h}[\Psi]&=S_{0}[\Psi]+h\omega(\Psi,\hat{m}_{0})\\
    &=\frac{1}{2}\omega(\Psi,m_{1}(\Psi))+\frac{1}{3}\omega(\Psi,m_{2}(\Psi,\Psi))+h\omega(\Psi,\hat{m}_{0}),
\end{align}
where $h$ is open-closed coupling constant \cite{Zwiebach:1992bw}. For later convenience, we introduce $m_{0}$ defined as
\begin{equation}
    m_{0}=h\hat{m}_{0}.
\end{equation}
This action has weak $A_{\infty}$ structure and the following conditions hold.
\begin{align}
    0&=m_{1}\qty(m_{0})\\
    0&=m_{1}\qty(m_{1}\qty(\Psi_{1}))+m_{2}\qty(m_{0},\Psi_{1})+(-1)^\abs{\Psi_{1}}m_{2}\qty(\Psi_{1},m_{0})\\
    0&=m_{1}\qty(m_{2}\qty(\Psi_{1},\Psi_{2}))+m_{2}\qty(m_{1}\qty(\Psi_{1}),\Psi_{2})+(-1)^\abs{\Psi_{1}}m_{2}\qty(\Psi_{1},m_{1}\qty(\Psi_{2}))\\
    0&=m_{2}\qty(m_{2}\qty(\Psi_{1},\Psi_{2}),\Psi_{3})+(-1)^\abs{\Psi_{1}}m_{2}\qty(\Psi_{1},m_{2}\qty(\Psi_{2},\Psi_{3}))
\end{align}
More precisely, because $m_{1},m_{2}$ satisfy \eqref{Ainf:1},\eqref{Ainf:2} and \eqref{Ainf:3}, new conditions are only the following two conditions.
\begin{align}
    0&=m_{1}\qty(m_{0})\\
    0&=m_{2}\qty(m_{0},\Psi_{1})+(-1)^\abs{\Psi_{1}}m_{2}\qty(\Psi_{1},m_{0})\label{wAinf:2}
\end{align}
It follows immediately from \eqref{cyclic} that $m_{0}$ is cyclic. Accordingly, we define the coderivation $\vb{m}^{h}$ as
\begin{equation}
    \vb{m}^{h}\coloneqq\vb{m}+\vb{m}_{0}.
\end{equation}
In the coalgebra language, the weak $A_{\infty}$ relations together with the cyclicity conditions can be expressed as
\begin{equation}
    (\vb{m}^{h})^2=0\qc\bra{\omega}\pi_{2}\vb{m}^{h}=0,
\end{equation}
and the action can be written as
\begin{equation}
    S_{h}[\Psi]=\int_{0}^{1}\dd{t}\bra{\omega}\pi_{1}\vb*{\partial}_t\frac{1}{1-\Psi(t)}\otimes\pi_{1}\vb{m}^{h}\frac{1}{1-\Psi(t)},
\end{equation}
where $\Psi(1)=\Psi,\Psi(0)=0$.

In this paper, we aim to find a field redefinition relating Witten's action to the deformed action. To do this, we consider the following field redefinition,
\begin{align}
    \Psi&=F[\Psi']\\
    &=F_{0}+F_{1}(\Psi')+F_{2}(\Psi',\Psi')+\dots,
\end{align}
where $F_{k}$ is multilinear map. In the coalgebra language, the field redefinition is described by a cohomomorphism $\vb{F}$.
\begin{equation}
    \frac{1}{1-\Psi}=\vb{F}\frac{1}{1-\Psi'}
\end{equation}
In particular, given coderivations $\vb{m}^{h}$ and $\vb{m}'$ and a cohomomorphism $\vb{F}$ satisfying
\begin{equation}
    \vb{m}^{h}\vb{F}=\vb{F}\vb{m}'\qc\bra{\omega}\vb{F}\otimes\vb{F}=\bra{\omega'},\label{cohomo}
\end{equation}
the action transforms as
\begin{equation}
    \int_{0}^{1}\dd{t}\bra{\omega}\pi_{1}\vb*{\partial}_t\frac{1}{1-\Psi(t)}\otimes\pi_{1}\vb{m}\frac{1}{1-\Psi(t)}=\int_{0}^{1}\dd{t}\bra{\omega'}\pi_{1}\vb*{\partial}_t\frac{1}{1-\Psi'(t)}\otimes\pi_{1}\vb{m}'\frac{1}{1-\Psi'(t)}.
\end{equation}
Here we must be mindful of the boundary conditions for $\Psi(t)$ and $\Psi'(t)$. Consistency of the field redefinition requires that
\begin{align}
    \frac{1}{1-\Psi(1)}&=\vb{F}\frac{1}{1-\Psi'(1)},\\
    \frac{1}{1-\Psi(0)}&=\vb{F}\frac{1}{1-\Psi'(0)}.
\end{align}
If we set $\Psi'(0)=0$, the boundary condition implies that $\Psi(0)$ must satisfy
\begin{equation}
    \frac{1}{1-\Psi(0)}=\frac{1}{1-F_{0}}.
\end{equation}
Therefore, when $\Psi'(0)=0$, we cannot impose $\Psi(0)=0$ \footnote{The author thanks Keisuke Konosu for discussion on this point.}. For this reason, we set $\Psi(0)=F_{0}$. With this choice, the action is shifted by a field-independent term evaluated on the background field $F_{0}$.
\begin{equation}
    \frac{1}{2}\omega(\Psi,m_{1}(\Psi))+\frac{1}{3}\omega(\Psi,m_{2}(\Psi,\Psi))+\omega(\Psi,m_{0})-S_{h}[F_{0}]=\sum_{n\ge0}\frac{1}{n+1}\omega'(\Psi',m'_{n}(\Psi',\dots,\Psi'))
\end{equation}

Next we assume that the field redefinition can be written as
\begin{equation}
    \vb{F}=e^{\vb*{\mu}},
\end{equation}
where $\vb*{\mu}$ is a coderivation. Because the cohomomorphism satisfies \eqref{cohomo}, the coderivation $\vb{m}'$ can be written as
\begin{align}
    \vb{m}'&=e^{-\vb*{\mu}}\vb{m}^{h}e^{\vb*{\mu}}\\
    &=\vb{m}^{h}-\comm{\vb*{\mu}}{\vb{m}^{h}}+\frac{1}{2}\comm{\vb*{\mu}}{\comm{\vb*{\mu}}{\vb{m}^{h}}}+\dots.
\end{align}
For later use, we recall that the graded commutator of two coderivations is again a coderivation. Given two coderivations $\vb{C}_{n}$ and $\vb{D}_{m}$ associated with multilinear maps $C_{n}$ and $D_{m}$, their commutator $\comm{\vb{C}_{n}}{\vb{D}_{m}}$ is explicitly given by
\begin{equation}
    \comm{\vb{C}_{n}}{\vb{D}_m}=\sum_{k=0}^\infty\qty(\sum_{l=0}^k\mathbb{I}^{\otimes k-l}\otimes\comm{C_{n}}{D_{m}}\otimes\mathbb{I}^{\otimes l})\pi_{k+n+m-1},
\end{equation}
where $\comm{C_{n}}{D_{m}}$ is defined by
\begin{equation}
    \comm{C_{n}}{D_{m}}=C_{n}\qty(\sum_{l=0}^{n-1}\mathbb{I}^{\otimes n-1-l}\otimes D_{m}\otimes\mathbb{I}^{\otimes l})-D_{m}\qty(\sum_{l=0}^{m-1}\mathbb{I}^{\otimes m-1-l}\otimes C_{n}\otimes\mathbb{I}^{\otimes l}).
\end{equation}
Moreover, if the coderivation $\vb*{\mu}$ is cyclic,
\begin{equation}
    \bra{\omega}\pi_{2}\vb*{\mu}=0,
\end{equation}
the symplectic inner product is invariant under the field redefinition,
\begin{equation}
    \bra{\omega}\pi_{2}\vb{F}=\bra{\omega}\pi_{2}.
\end{equation}
On the other hand, if the coderivation $\vb*{\mu}$ is not cyclic but instead satisfies
\begin{equation}
    \bra{\omega}\pi_{2}\vb*{\mu}=\alpha\bra{\omega}\pi_{2},
\end{equation}
for a constant $\alpha$, the symplectic inner product is rescaled by a constant factor $e^{\alpha}$,
\begin{equation}
    \bra{\omega}\pi_{2}\vb{F}=e^\alpha\bra{\omega}\pi_{2}.
\end{equation}
The simplest example of this is a constant rescaling of the string field
\begin{equation}
    \Psi=\frac{1}{g}\Psi'.
\end{equation}
We already know that by this redefinition the action changes as
\begin{align}
    &\frac{1}{2}\omega(\Psi,m_{1}(\Psi))+\frac{g}{3}\omega(\Psi,\hat{m}_{2}(\Psi,\Psi))+\frac{1}{g}\omega(\Psi,\hat{m}_{0})\\
    &=\frac{1}{g^2}\qty(\frac{1}{2}\omega(\Psi',m_{1}(\Psi'))+\frac{1}{3}\omega(\Psi',\hat{m}_{2}(\Psi',\Psi'))+\omega(\Psi',\hat{m}_{0})),
\end{align}
where we write $m_{2}=g\hat{m}_{2}$ and set $h$ to $g^{-1}$. Let us rephrase this example in the language of homotopy algebra. In coalgebra language, this amounts to finding a cohomomorphism $\vb{F}$ such that
\begin{align}
    \qty(\vb{m}_{1}+g\hat{\vb{m}}_{2}+\frac{1}{g}\hat{\vb{m}}_{0})\vb{F}&=\vb{F}(\vb{m}_{1}+\hat{\vb{m}}_{2}+\hat{\vb{m}}_{0}),\\
    \bra{\omega}\vb{F}\otimes\vb{F}&=\frac{1}{g^{2}}\bra{\omega},\label{ex:Fcond}
\end{align}
or, equivalently, a coderivation $\vb*{\mu}$ satisfying
\begin{align}
    (\vb{m}_{1}+\hat{\vb{m}}_{2}+\hat{\vb{m}}_{0})&=e^{-\vb*{\mu}}\qty(\vb{m}_{1}+g\hat{\vb{m}}_{2}+\frac{1}{g}\hat{\vb{m}}_{0})e^{\vb*{\mu}},\\
    \bra{\omega}\pi_{2}\vb*{\mu}&=-2\ln g\bra{\omega}\pi_{2}.\label{ex:mucond}
\end{align}
One such coderivation $\vb*{\mu}$ is given by
\begin{equation}
    \vb*{\mu}=-(\ln g)\vb{I}\qc\vb{I}\coloneqq\sum_{k=0}^\infty\qty(\sum_{l=0}^k\mathbb{I}^{\otimes k-l}\otimes\mathbb{I}\otimes\mathbb{I}^{\otimes l})\pi_{k+1}=\sum_{k=0}^\infty(k+1)\mathbb{I}^{\otimes k+1}\pi_{k+1},
\end{equation}
where $\vb{I}$ is defined by the identity map $\mathbb{I}$ on $\mathcal{H}$. The field redefinition then reads
\begin{align}
    \pi_{1}\vb{F}\frac{1}{1-\Psi'}&=\Psi'+\mu(\Psi')+\frac{1}{2}\mu(\mu(\Psi'))+\dots\\
    &=\frac{1}{g}\Psi'.
\end{align}
Additionally, since the coderivation $\vb*{\mu}$ satisfies \eqref{ex:mucond}, the symplectic form transforms by an overall factor such as \eqref{ex:Fcond}

Note that $\vb{I}$ is not the identity operator on the tensor algebra $T\mathcal{H}$.
In particular, for any multilinear map $C_{n}$, we find
\begin{align}
    \comm{C_{n}}{\mathbb{I}}&=C_{n}\qty(\sum_{l=0}^{n-1}\mathbb{I}^{\otimes n-1-l}\otimes\mathbb{I}\otimes\mathbb{I}^{\otimes l})-\mathbb{I}\qty(\sum_{l=0}^0\mathbb{I}^{\otimes -l}\otimes C_{n}\otimes\mathbb{I}^{\otimes l})\\
    &=(n-1)C_{n},
\end{align}
and consequently,
\begin{equation}
    \comm{\vb{C}_{n}}{\vb{I}}=(n-1)\vb{C}_{n}.
\end{equation}
Using this relation, we obtain
\begin{align}
    e^{-\vb*{\mu}}\vb{C}_{n}e^{\vb*{\mu}}&=\vb{C}_{n}-\comm{\vb*{\mu}}{\vb{C}_{n}}+\frac{1}{2}\comm{\vb*{\mu}}{\comm{\vb*{\mu}}{\vb{C}_{n}}}+\dots\\
    &=\vb{C}_{n}+\ln g\comm{\vb{I}}{\vb{C}_{n}}+\frac{1}{2}\qty(\ln g)^{2}\comm{\vb{I}}{\comm{\vb{I}}{\vb{C}_{n}}}+\dots\\
    &=g^{1-n}\vb{C}_{n}.
\end{align}
This demonstrates that the above field redefinition correctly reproduces the expected rescaling behavior.

In the next section, we look for a field redefinition relating the deformed action to Witten's action. This is equivalent to finding a coderivation $\vb*{\mu}$ such that
\begin{align}
    \vb{m}^{h}&=e^{-\vb*{\mu}}\vb{m}e^{\vb*{\mu}},\label{cond:frdc}\\
    \bra{\omega}\pi_{2}\vb*{\mu}&=0.
\end{align}
To get more detailed insight, we expand the coderivation $\vb*{\mu}$ as a power series in the open-closed coupling constant $h$,
\begin{equation}
    \vb*{\mu}=\sum_{l\ge0}h^l\vb*{\mu}^{(l)}\qc\vb*{\mu}^{(l)}\coloneqq\sum_{k\ge0}\vb*{\mu}_{k}^{(l)}.
\end{equation}
Expanding \eqref{cond:frdc} order by order in $h$, we obtain
\begin{align}
    0&=\comm{\vb*{\mu}^{(0)}}{\vb{m}}+\frac{1}{2}\comm{\vb*{\mu}^{(0)}}{\comm{\vb*{\mu}^{(0)}}{\vb{m}}}+\dots,\label{cond:frdc0}\\
    \hat{\vb{m}}_{0}&=\comm{\vb*{\mu}^{(1)}}{\vb{m}}+\frac{1}{2}\comm{\vb*{\mu}^{(1)}}{\comm{\vb*{\mu}^{(0)}}{\vb{m}}}+\frac{1}{2}\comm{\vb*{\mu}^{(0)}}{\comm{\vb*{\mu}^{(1)}}{\vb{m}}}+\dots,\\
    &\vdots\\
    0&=\sum_{n=0}^\infty\frac{1}{n!}\sum_{i_{1}+\dots+i_{n}=k}[\vb*{\mu}^{(i_{1})},[\dots,[\vb*{\mu}^{(i_{n})},\vb{m}]]\dots]\qfor k\ge2.
\end{align}

Before closing this section, we comment on the relation between our field redefinition to be constructed in this work and that discussed in \cite{Baba:2012cs}. In the earlier work, it was shown that the deformed action in a weak gravitational background is related to Witten's action through a field redefinition as
\begin{equation}
    S_{h}[\Psi]=(1+h)S_{0}[\Psi']+\order{h^{2}}.\label{relation:bi}
\end{equation}
More precisely, by setting $\vb*{\mu}^{(0)}=0$ and working up to order $h$, the authors constructed a coderivation satisfying
\begin{equation}
    \vb{m}_{0}=\comm{h\vb*{\mu}^{(1)}_{0}+h\vb*{\mu}^{(1)}_{1}}{\vb{m}}\qc\bra{\omega}\pi_{2}\vb*{\mu}^{(1)}_{1}=\bra{\omega}\pi_{2}.
\end{equation}
When expressed in terms of the multilinear maps, these conditions take the form\footnote{In \cite{Baba:2012cs}, $\mu^{(1)}_{0}$ and $\mu^{(1)}_{1}$ are denoted by $\chi$ and $\mathcal{G}$, respectively.}
\begin{align}
    \hat{m}_{0}&=-m_{1}\qty(\mu^{(1)}_{0}),\\
    0&=\mu^{(1)}_{1}\qty(m_{1}\qty(\Psi_{1}))-m_{1}\qty(\mu^{(1)}_{1}(\Psi))-m_{2}\qty(\mu^{(1)}_{0},\Psi)-m_{2}\qty(\Psi,\mu^{(1)}_{0}),\\
    0&=\mu^{(1)}_{1}\qty(m_{2}\qty(\Psi_{1},\Psi_{2}))-m_{2}\qty(\mu^{(1)}_{1}(\Psi_{1}),\Psi_{2})-m_{2}\qty(\Psi_{1},\mu^{(1)}_{1}(\Psi_{2})),\\
    \omega(\Psi_{1},\Psi_{2})&=\omega(\mu^{(1)}_{1}(\Psi_{1}),\Psi_{2})+\omega(\Psi_{1},\mu^{(1)}_{1}(\Psi_{2})).
\end{align}
At first sight, this may appear to be in contradiction with our aim of finding a field redefinition satisfying
\begin{equation}
    S_{h}[\Psi]=S_{0}[\Psi'],
\end{equation}
but the two results are compatible. First, in the earlier work, the multilinear maps $\mu^{(1)}_{0}$ and $\mu^{(1)}_{1}$ were constructed using singular operators. In fact, for certain string fields, $\mu^{(1)}_{0}$ and $\mu^{(1)}_{1}$ do not satisfy the required conditions \cite{Hata:2013hba,Ando:2023dhl}. In particular, according to \cite{Ando:2023dhl}, for string fields $\Psi_{1}$ and $\Psi_{2}$ constructed from the $K,B,c$ and matter operators involving $X^{0}$, the following two conditions fail to hold, at least at the numerical level.
\begin{align}
    \omega(\Psi_{1},\hat{m}_{0})&=-\omega(\Psi_{1},m_{1}(\mu^{(1)}_{0}))\\
    0&=\omega(\Psi_{1},\mu^{(1)}_{1}(m_{1}(\Psi_{2})))-\omega(\Psi_{1},m_{1}(\mu^{(1)}_{1}(\Psi_{2})))\\
    &\qquad-\omega(\Psi_{1},m_{2}(\mu^{(1)}_{0},\Psi_{2}))-\omega(\Psi_{1},m_{2}(\Psi_{2},\mu^{(1)}_{0}))
\end{align}
Therefore, even in a weak gravitational background, it is not clear whether the relation \eqref{relation:bi} is satisfied. On the other hand, as will be seen in the next section, the multilinear maps constructed in this paper do not involve singular operators. Nevertheless, when they are expressed in the star-product notation, we will find in section \ref{star} that the resulting field redefinition is still not well-defined. In this sense, our construction is compatible with the observations in the earlier work.

\section{Field redefinition between Witten's theory and deformed theory\label{fieldredef}}
In this section, we construct appropriate multilinear maps and determine a field redefinition relating Witten's theory to the deformed theory.

Our task is to find coderivations that satisfy the conditions derived in the previous section. The simplest solution to \eqref{cond:frdc0} is to set $\vb*{\mu}^{(0)}=0$. With this choice, the conditions reduce to
\begin{align}
    \hat{\vb{m}}_{0}&=\comm{\vb*{\mu}^{(1)}}{\vb{m}},\label{cond:redef1}\\
    0&=\comm{\vb*{\mu}^{(2)}}{\vb{m}}+\frac{1}{2}\comm{\vb*{\mu}^{(1)}}{\comm{\vb*{\mu}^{(1)}}{\vb{m}}},\label{cond:redef2}\\
    &\vdots
\end{align}
Moreover, expanding \eqref{cond:redef1} in terms of the component multilinear maps, we obtain
\begin{align}
    \hat{m}_{0}&=-m_{1}\qty(\mu^{(1)}_{0}),\label{cond:0}\\
    0&=\mu^{(1)}_{1}\qty(m_{1}\qty(\Psi))-m_{1}\qty(\mu^{(1)}_{1}\qty(\Psi))-m_{2}\qty(\mu^{(1)}_{0},\Psi)-m_{2}\qty(\Psi,\mu^{(1)}_{0}),\\
    0&=\mu^{(1)}_{1}\qty(m_{2}\qty(\Psi_{1},\Psi_{2}))-m_{2}\qty(\mu^{(1)}_{1}\qty(\Psi_{1}),\Psi_{2})-m_{2}\qty(\Psi_{1},\mu^{(1)}_{1}\qty(\Psi_{2}))\\
    &\qquad+\mu^{(1)}_{2}\qty(m_{1}\qty(\Psi_{1}),\Psi_{2})+\mu^{(1)}_{2}\qty(\Psi_{1},m_{1}\qty(\Psi_{2}))-m_{1}\qty(\mu^{(1)}_{2}\qty(\Psi_{1},\Psi_{2})),\\
    0&=\mu^{(1)}_{2}\qty(m_{2}\qty(\Psi_{1},\Psi_{2}),\Psi_{3})+\mu^{(1)}_{2}\qty(\Psi_{1},m_{2}\qty(\Psi_{2},\Psi_{3}))-m_{2}\qty(\mu^{(1)}_{2}\qty(\Psi_{1},\Psi_{2}),\Psi_{3})-m_{2}\qty(\Psi_{1},\mu^{(1)}_{2}\qty(\Psi_{2},\Psi_{3}))\\
    &\qquad+\mu^{(1)}_{3}\qty(m_{1}\qty(\Psi_{1}),\Psi_{2},\Psi_{3})+\mu^{(1)}_{3}\qty(\Psi_{1},m_{1}\qty(\Psi_{2}),\Psi_{3})+\mu^{(1)}_{3}\qty(\Psi_{1},\Psi_{2},m_{1}\qty(\Psi_{3}))-m_{1}\qty(\mu^{(1)}_{3}(\Psi_{1},\Psi_{2},\Psi_{3})),\\
    &\vdots.
\end{align}
However, according to \eqref{cond:0}, the on-shell closed string state must be BRS exact. Since all amplitudes involving $m_{0}$ vanish in this case, such an $m_{0}$ does not contribute to any physical effect. In order to obtain a nontrivial deformation, we consider Witten's theory around a tachyon vacuum solution $\Psi_{\mathrm{tv}}$ and its action is given by
\begin{align}
    S^{\mathrm{tv}}[\Psi]&\coloneqq S[\Psi_{\mathrm{tv}}+\Psi]-S[\Psi_{\mathrm{tv}}]\\
    &=\frac{1}{2}\omega(\Psi,m^{\mathrm{tv}}_{1}(\Psi))+\frac{1}{3}\omega(\Psi,m_{2}(\Psi,\Psi)),
\end{align}
where $m^{\mathrm{tv}}_{1}$ is defined as
\begin{equation}
    m^{\mathrm{tv}}_{1}(\Psi)\equiv m_{1}(\Psi)+m_{2}(\Psi_{\mathrm{tv}},\Psi)+(-1)^\abs{\Psi}m_{2}(\Psi,\Psi_{\mathrm{tv}}).
\end{equation}
Similarly, the deformed action around $\Psi_{\mathrm{tv}}$ is given by
\begin{align}
    S^{\mathrm{tv}}_{h}[\Psi]&\coloneqq S_{h}[\Psi_{\mathrm{tv}}+\Psi]-S_{h}[\Psi_{\mathrm{tv}}]\\
    &=\frac{1}{2}\omega(\Psi,m^{\mathrm{tv}}_{1}(\Psi))+\frac{1}{3}\omega(\Psi,m_{2}(\Psi,\Psi))+h\omega(\Psi,\hat{m}_{0}).
\end{align}
Note that the tachyon vacuum solution $\Psi_{\mathrm{tv}}$ is a classical solution of Witten's theory, but not of the deformed theory. Indeed, $\Psi_{\mathrm{tv}}$ satisfies
\begin{equation}
    m_{1}(\Psi_{\mathrm{tv}})+m_{2}(\Psi_{\mathrm{tv}},\Psi_{\mathrm{tv}})=0,
\end{equation}
but it does not satisfy
\begin{equation}
    m_{1}(\Psi_{\mathrm{tv}})+m_{2}(\Psi_{\mathrm{tv}},\Psi_{\mathrm{tv}})+m_{0}=0.
\end{equation}
Since this action has a weak $A_{\infty}$ structure, we define the corresponding coderivations by
\begin{equation}
    \vb{m}^{\mathrm{tv}}\coloneqq\vb{m}^{\mathrm{tv}}_{1}+\vb{m}_{2}\qc\vb{m}^{h,\mathrm{tv}}\coloneqq\vb{m}^{\mathrm{tv}}+\vb{m}_{0}.
\end{equation}
In complete analogy with the case of the action around the perturbative vacuum, these coderivations satisfy
\begin{alignat}{2}
    (\vb{m}^{\mathrm{tv}})^2&=0\qc&\bra{\omega}\pi_{2}\vb{m}^{\mathrm{tv}}&=0,\\
    (\vb{m}^{h,\mathrm{tv}})^2&=0\qc&\bra{\omega}\pi_{2}\vb{m}^{h,\mathrm{tv}}&=0.
\end{alignat}
We then look for cohomomorphism $\vb{F}$ such that
\begin{equation}
    \vb{m}^{h,\mathrm{tv}}\vb{F}=\vb{F}\vb{m}^{\mathrm{tv}}.
\end{equation}
Assuming $\vb{F}=e^{\vb*{\mu}}$ and setting $\vb*{\mu}^{(0)}=0$, the analysis proceeds exactly as before, with the replacement $\vb{m}^{h}\to\vb{m}^{h,\mathrm{tv}}$ and $\vb{m}\to\vb{m}^{\mathrm{tv}}$. We then obtain
\begin{align}
    \hat{\vb{m}}_{0}&=\comm{\vb*{\mu}^{(1)}}{\vb{m}^{\mathrm{tv}}},\label{cond}\\
    0&=\comm{\vb*{\mu}^{(2)}}{\vb{m}^{\mathrm{tv}}}+\frac{1}{2}\comm{\vb*{\mu}^{(1)}}{\comm{\vb*{\mu}^{(1)}}{\vb{m}^{\mathrm{tv}}}},\label{condhigher}\\
    &\vdots,
\end{align}
together with the reduced cyclicity condition
\begin{equation}
    \bra{\omega}\pi_{2}\vb*{\mu}=0.\label{redcyc}
\end{equation}

An advantage of considering the action around the tachyon vacuum solution is the existence of a homotopy operator $A$ satisfying
\begin{equation}
    m^{\mathrm{tv}}_{1}(A)=1,\label{def:homotopy}
\end{equation}
where 1 denotes identity string field, and the degrees of $A$ and $1$ are given by $\abs{A}=-2$ and $\abs{1}=-1$, respectively. For any string field $\Psi$, the identity string field obeys
\begin{equation}
    -m_{2}(1,\Psi)=(-1)^\abs{\Psi}m_{2}(\Psi,1)=\Psi.\label{def:identitystring}
\end{equation}
At first sight, this relation, together with the degree assignments of $A$ and $1$, may appear unusual. In Appendix \ref{homotopyop}, we briefly verify that these conventions are consistent with the star product notation. Using only the cyclic $A_{\infty}$ relations, the defining property of the homotopy operator \eqref{def:homotopy}, and the definition of the identity string field \eqref{def:identitystring}, we will construct a field redefinition for which the conditions \eqref{cond} and \eqref{condhigher} are satisfied.

First, by expanding \eqref{cond} into component products, we obtain
\begin{align}
    \hat{\vb{m}}_{0}&=\comm{\vb*{\mu}^{(1)}_{0}}{\vb{m}^{\mathrm{tv}}_{1}},\\
    0&=\comm{\vb*{\mu}^{(1)}_{n}}{\vb{m}^{\mathrm{tv}}_{1}}+\comm{\vb*{\mu}^{(1)}_{n-1}}{\vb{m}_{2}}\qfor n\ge1,
\end{align}
or equivalently
\begin{align}
    \hat{m}_{0}&=-m^{\mathrm{tv}}_{1}\qty(\mu^{(1)}_{0}),\label{cond:1}\\
    0&=\mu^{(1)}_{1}\qty(m^{\mathrm{tv}}_{1}(\Psi))-m^{\mathrm{tv}}_{1}\qty(\mu^{(1)}_{1}(\Psi))-m_{2}\qty(\mu^{(1)}_{0},\Psi)-m_{2}\qty(\Psi,\mu^{(1)}_{0}),\label{cond:2}\\
    0&=\mu^{(1)}_{1}\qty(m_{2}(\Psi_{1},\Psi_{2}))-m_{2}\qty(\mu^{(1)}_{1}(\Psi_{1}),\Psi_{2})-m_{2}\qty(\Psi_{1},\mu^{(1)}_{1}(\Psi_{2}))\\
    &\qquad+\mu^{(1)}_{2}\qty(m^{\mathrm{tv}}_{1}(\Psi_{1}),\Psi_{2})+\mu^{(1)}_{2}\qty(\Psi_{1},m^{\mathrm{tv}}_{1}(\Psi_{2}))-m^{\mathrm{tv}}_{1}\qty(\mu^{(1)}_{2}\qty(\Psi_{1},\Psi_{2})),\label{cond:3}\\
    0&=\mu^{(1)}_{2}\qty(m_{2}(\Psi_{1},\Psi_{2}),\Psi_{3})+\mu^{(1)}_{2}\qty(\Psi_{1},m_{2}(\Psi_{2},\Psi_{3}))-m_{2}\qty(\mu^{(1)}_{2}(\Psi_{1},\Psi_{2}),\Psi_{3})-m_{2}\qty(\Psi_{1},\mu^{(1)}_{2}(\Psi_{2},\Psi_{3}))\\
    &\quad+\mu^{(1)}_{3}\qty(m^{\mathrm{tv}}_{1}(\Psi_{1}),\Psi_{2},\Psi_{3})+\mu^{(1)}_{3}\qty(\Psi_{1},m^{\mathrm{tv}}_{1}(\Psi_{2}),\Psi_{3})+\mu^{(1)}_{3}\qty(\Psi_{1},\Psi_{2},m^{\mathrm{tv}}_{1}(\Psi_{3}))-m_{1}^{\mathrm{tv}}\qty(\mu^{(1)}_{3}(\Psi_{1},\Psi_{2},\Psi_{3})),\\
    &\vdots.
\end{align}
We now define $\mu^{(1)}_{0}$ by
\begin{equation}
    \mu^{(1)}_{0}\coloneqq m_{2}(\hat{m}_{0},A)=-m_{2}(A,\hat{m}_{0}).
\end{equation}
Using only the $A_{\infty}$ relation \eqref{Ainf:2}, the defining property of the homotopy operator \eqref{def:homotopy}, and the identity string field relation \eqref{def:identitystring}, it is straightforward to verify that this definition satisfies \eqref{cond:1}.
\begin{align}
    -m^{\mathrm{tv}}_{1}(m_{2}(\hat{m}_{0},A))&=m_{2}\qty(m^{\mathrm{tv}}_{1}(\hat{m}_{0}),A)-m_{2}\qty(\hat{m}_{0},m^{\mathrm{tv}}_{1}(A))\\
    &=-m_{2}(\hat{m}_{0},1)\\
    &=\hat{m}_{0}
\end{align}

The next step is to construct $\mu^{(1)}_{1}$ so that \eqref{cond:2} is satisfied. We define $\mu^{(1)}_{1}$ by
\begin{equation}
    \mu^{(1)}_{1}(\Psi)\coloneqq m_{2}(m_{2}(\mu^{(1)}_{0},\Psi),A)=-m_{2}(A,m_{2}(\Psi,\mu^{(1)}_{0})),
\end{equation}
where the second equality is follows from the identity
\begin{align}
    m_{2}(m_{2}(\mu^{(1)}_{0},\Psi),A)&=-m_{2}(m_{2}(m_{2}(A,\hat{m}_{0}),\Psi),A)\\
    &=m_{2}(m_{2}(A,m_{2}(\hat{m}_{0},\Psi)),A)\\
    &=-(-1)^\abs{\Psi}m_{2}(m_{2}(A,m_{2}(\Psi,\hat{m}_{0})),A)\\
    &=(-1)^\abs{\Psi}m_{2}(A,m_{2}(m_{2}(\Psi,\hat{m}_{0}),A))\\
    &=-m_{2}(A,m_{2}(\Psi,m_{2}(\hat{m}_{0},A)))\\
    &=-m_{2}(A,m_{2}(\Psi,\mu^{(1)}_{0})),\label{cyc:mu1}
\end{align}
which is a consequence of \eqref{Ainf:3} and \eqref{wAinf:2}. Using only the $A_{\infty}$ relations \eqref{Ainf:2} and \eqref{Ainf:3}, the weak $A_{\infty}$ relation \eqref{wAinf:2}, the defining property of the homotopy operator \eqref{def:homotopy}, and the identity string field relation \eqref{def:identitystring}, we compute
\begin{align}
    m^{\mathrm{tv}}_{1}\qty(\mu^{(1)}_{1}\qty(\Psi))&=m^{\mathrm{tv}}_{1}\qty(m_{2}(m_{2}(\mu^{(1)}_{0},\Psi),A))\\
    &=-m_{2}(m^{\mathrm{tv}}_{1}(m_{2}(\mu^{(1)}_{0},\Psi)),A)-m_{2}(\mu^{(1)}_{0},\Psi)\\
    &=m_{2}(m_{2}(m^{\mathrm{tv}}_{1}(\mu^{(1)}_{0}),\Psi),A)+m_{2}(m_{2}(\mu^{(1)}_{0},m^{\mathrm{tv}}_{1}(\Psi)),A)-m_{2}(\mu^{(1)}_{0},\Psi)\\
    &=-m_{2}(m_{2}(\hat{m}_{0},\Psi),A)+\mu^{(1)}_{1}(m^{\mathrm{tv}}_{1}(\Psi))-m_{2}(\mu^{(1)}_{0},\Psi)\\
    &=(-1)^\abs{\Psi}m_{2}(m_{2}(\Psi,\hat{m}_{0}),A)+\mu^{(1)}_{1}(m^{\mathrm{tv}}_{1}(\Psi))-m_{2}(\mu^{(1)}_{0},\Psi)\\
    &=-m_{2}(\Psi,m_{2}(\hat{m}_{0},A))+\mu^{(1)}_{1}(m^{\mathrm{tv}}_{1}(\Psi))-m_{2}(\mu^{(1)}_{0},\Psi)\\
    &=-m_{2}(\Psi,\mu^{(1)}_{0})+\mu^{(1)}_{1}(m^{\mathrm{tv}}_{1}(\Psi))-m_{2}(\mu^{(1)}_{0},\Psi).
\end{align}
This shows that \eqref{cond:2} is indeed satisfied.

The remaining multilinear maps $\mu^{(1)}_{k}$ for $k\ge2$ can be defined inductively. Suppose that, for some $k\ge1$, the maps $\mu^{(1)}_{k}$ and $\mu^{(1)}_{k-1}$ are already constructed and satisfy
\begin{equation}
    0=\comm{\mu^{(1)}_{k}}{m^{\mathrm{tv}}_{1}}+\comm{\mu^{(1)}_{k-1}}{m_{2}}.
\end{equation}
We then define $\mu^{(1)}_{k+1}$ by
\begin{equation}
    \mu^{(1)}_{k+1}(\Psi_{1},\dots,\Psi_{k+1})\coloneqq m_{2}(m_{2}(\mu^{(1)}_{k}(\Psi_{1},\dots,\Psi_{k}),\Psi_{k+1}),A).
\end{equation}
We claim that $\mu^{(1)}_{k+1}$ and $\mu^{(1)}_{k}$ satisfy
\begin{equation}
    0=\comm{\mu^{(1)}_{k+1}}{m^{\mathrm{tv}}_{1}}+\comm{\mu^{(1)}_{k}}{m_{2}}.
\end{equation}
The proof proceeds as in the previous cases. Acting with $m^{\mathrm{tv}}_{1}$ on $\mu^{(1)}_{k+1}$, we obtain
\begin{align}
    &m^{\mathrm{tv}}_{1}\qty(\mu^{(1)}_{k+1}(\Psi_{1},\dots,\Psi_{k+1}))\\
    &=-m_{2}(m^{\mathrm{tv}}_{1}(m_{2}(\mu^{(1)}_{k}(\Psi_{1},\dots,\Psi_{k}),\Psi_{k+1})),A)-m_{2}(m_{2}(\mu^{(1)}_{k}(\Psi_{1},\dots,\Psi_{k}),\Psi_{k+1}))\\
    &=m_{2}(m_{2}(m^{\mathrm{tv}}_{1}(\mu^{(1)}_{k}(\Psi_{1},\dots,\Psi_{k})),\Psi_{k+1}),A)+(-1)^{\abs{\Psi_{1}}+\dots+\abs{\Psi_{k}}}m_{2}(m_{2}(\mu^{(1)}_{k}(\Psi_{1},\dots,\Psi_{k}),m^{\mathrm{tv}}_{1}(\Psi_{k+1})),A)\\
    &\qquad-m_{2}(\mu^{(1)}_{k}(\Psi_{1},\dots,\Psi_{k}),\Psi_{k+1})\\
    &=m_{2}(m_{2}(\mu^{(1)}_{k}(m^{\mathrm{tv}}_{1}(\Psi_{1}),\dots,\Psi_{k}),\Psi_{k+1}),A)+\dots\\
    &\qquad\quad\dots+(-1)^{\abs{\Psi_{1}}+\dots+\abs{\Psi_{n-1}}}m_{2}(m_{2}(\mu^{(1)}_{k}(\Psi_{1},\dots,m^{\mathrm{tv}}_{1}(\Psi_{k})),\Psi_{k+1}),A)\\
    &\qquad+m_{2}(m_{2}(\comm{m^{\mathrm{tv}}_{1}}{\mu^{(1)}_{k}}(\Psi_{1},\dots,\Psi_{k}),\Psi_{k+1}),A)\\
    &\qquad+(-1)^{\abs{\Psi_{1}}+\dots+\abs{\Psi_{k}}}m_{2}(m_{2}(\mu^{(1)}_{k}(\Psi_{1},\dots,\Psi_{k}),m^{\mathrm{tv}}_{1}(\Psi_{k+1})),A)-m_{2}(\mu^{(1)}_{k}(\Psi_{1},\dots,\Psi_{k}),\Psi_{k+1})\\
    &=\mu^{(1)}_{k+1}(m^{\mathrm{tv}}_{1}(\Psi_{1}),\dots,\Psi_{k},\Psi_{k+1})+\dots+(-1)^{\abs{\Psi_{1}}+\dots+\abs{\Psi_{k-1}}}\mu^{(1)}_{k+1}(\Psi_{1},\dots,m^{\mathrm{tv}}_{1}(\Psi_{k}),\Psi_{k+1})\\
    &\qquad-m_{2}(m_{2}(\comm{m_{2}}{\mu^{(1)}_{k-1}}(\Psi_{1},\dots,\Psi_{k}),\Psi_{k+1}),A)\\
    &\qquad+(-1)^{\abs{\Psi_{1}}+\dots+\abs{\Psi_{k}}}\mu^{(1)}_{k+1}(\Psi_{1},\dots,\Psi_{k},m^{\mathrm{tv}}_{1}(\Psi_{k+1}))-m_{2}(\mu^{(1)}_{k}(\Psi_{1},\dots,\Psi_{k}),\Psi_{k+1}).
\end{align}
Therefore, evaluating the graded commutator explicitly, we find
\begin{align}
    &\comm{\mu^{(1)}_{k+1}}{m^{\mathrm{tv}}_{1}}(\Psi_{1},\dots,\Psi_{k+1})\\
    &=m_{2}(m_{2}(\comm{m_{2}}{\mu^{(1)}_{k-1}}(\Psi_{1},\dots,\Psi_{k}),\Psi_{k+1}),A)+m_{2}(\mu^{(1)}_{k}(\Psi_{1},\dots,\Psi_{k}),\Psi_{k+1})\\
    &=m_{2}(m_{2}(m_{2}(\mu^{(1)}_{k-1}(\Psi_{1},\dots,\Psi_{k-1}),\Psi_{k}),\Psi_{k+1}),A)+m_{2}(m_{2}(m_{2}(\Psi_{1},\mu^{(1)}_{k-1}(\Psi_{2},\dots,\Psi_{k})),\Psi_{k+1}),A)\\
    &\qquad-m_{2}(m_{2}(\mu^{(1)}_{k-1}(m_{2}(\Psi_{1},\Psi_{2}),\dots,\Psi_{k}),\Psi_{k+1}),A)-\dots\\
    &\qquad\quad\dots-(-1)^{\abs{\Psi_{1}}+\dots+\abs{\Psi_{k-2}}}m_{2}(m_{2}(\mu^{(1)}_{k-1}(\Psi_{1},\dots,m_{2}(\Psi_{k-1},\Psi_{k})),\Psi_{k+1}),A)\\
    &\qquad+m_{2}(\mu^{(1)}_{k}(\Psi_{1},\dots,\Psi_{k}),\Psi_{k+1})\\
    &=-(-1)^{\abs{\Psi_{1}}+\dots+\abs{\Psi_{k-1}}}m_{2}(m_{2}(\mu^{(1)}_{k-1}(\Psi_{1},\dots,\Psi_{k-1}),m_{2}(\Psi_{k},\Psi_{k+1})),A)\\
    &\qquad+m_{2}(\Psi_{1},m_{2}(m_{2}(\mu^{(1)}_{k-1}(\Psi_{2},\dots,\Psi_{k}),\Psi_{k+1}),A))\\
    &\qquad-m_{2}(m_{2}(\mu^{(1)}_{k-1}(m_{2}(\Psi_{1},\Psi_{2}),\dots,\Psi_{k}),\Psi_{k+1}),A)-\dots\\
    &\qquad\quad\dots-(-1)^{\abs{\Psi_{1}}+\dots+\abs{\Psi_{n-2}}}m_{2}(m_{2}(\mu^{(1)}_{k-1}(\Psi_{1},\dots,m_{2}(\Psi_{k-1},\Psi_{k})),\Psi_{k+1}),A)\\
    &\qquad+m_{2}(\mu^{(1)}_{k}(\Psi_{1},\dots,\Psi_{k}),\Psi_{k+1})\\
    &=-(-1)^{\abs{\Psi_{1}}+\dots+\abs{\Psi_{k-1}}}\mu^{(1)}_{k}(\Psi_{1},\dots,\Psi_{k-1},m_{2}(\Psi_{k},\Psi_{k+1}))\\
    &\qquad+m_{2}(\Psi_{1},\mu^{(1)}_{k}(\Psi_{2},\dots,\Psi_{k},\Psi_{k+1}))\\
    &\qquad-\mu^{(1)}_{k}(m_{2}(\Psi_{1},\Psi_{2}),\dots,\Psi_{k+1})-\dots-(-1)^{\abs{\Psi_{1}}+\dots+\abs{\Psi_{n-2}}}\mu^{(1)}_{k}(\Psi_{1},\dots,m_{2}(\Psi_{k-1},\Psi_{k}),\Psi_{k+1})\\
    &\qquad+m_{2}(\mu^{(1)}_{k}(\Psi_{1},\dots,\Psi_{k}),\Psi_{k+1})\\
    &=-\comm{\mu^{(1)}_{k}}{m_{2}}(\Psi_{1},\dots,\Psi_{k+1}).
\end{align}
This is what we want to show. Consequently, the multilinear maps
\begin{align}
    \mu^{(1)}_{0}&=m_{2}(\hat{m}_{0},A),\\
    \mu^{(1)}_{k+1}&=m_{2}(m_{2}(\mu^{(1)}_{k}(\Psi_{1},\dots,\Psi_{k}),\Psi_{k+1}),A)\qfor k\ge0
\end{align}
satisfy \eqref{cond}.

For our purposes, this is still not sufficient, and we must also verify the cyclicity condition \eqref{redcyc}. Before turning to this condition, we first show that the identity
\begin{equation}
    m_{2}(m_{2}(\mu^{(1)}_{k}(\Psi_{1},\dots,\Psi_{k}),\Psi_{k+1}),A)=-m_{2}(A,m_{2}(\Psi_{1},\mu^{(1)}_{k}(\Psi_{2},\dots,\Psi_{k+1}))),\label{mucyc}
\end{equation}
holds provided that
\begin{equation}
    m_{2}(m_{2}(\mu^{(1)}_{k-1}(\Psi_{1},\dots,\Psi_{k-1}),\Psi_{k}),A)=-m_{2}(A,m_{2}(\Psi_{1},\mu^{(1)}_{k-1}(\Psi_{2},\dots,\Psi_{k})))
\end{equation}
is satisfied. This follows directly from the recursive definition of $\mu^{(1)}_{k}$. Indeed, we compute
\begin{align}
    m_{2}(m_{2}(\mu^{(1)}_{k}(\Psi_{1},\dots,\Psi_{k}),\Psi_{k+1}),A)&=m_{2}(m_{2}(m_{2}(m_{2}(\mu^{(1)}_{k-1}(\Psi_{1},\dots,\Psi_{k-1}),\Psi_{k}),A),\Psi_{k+1}),A)\\
    &=-m_{2}(m_{2}(m_{2}(A,m_{2}(\Psi_{1},\mu^{(1)}_{k-1}(\Psi_{2},\dots,\Psi_{k}))),\Psi_{k+1}),A)\\
    &=-m_{2}(A,m_{2}(m_{2}(m_{2}(\Psi_{1},\mu^{(1)}_{k-1}(\Psi_{2},\dots,\Psi_{k})),\Psi_{k+1}),A))\\
    &=-m_{2}(A,m_{2}(\Psi_{1},m_{2}(m_{2}(\mu^{(1)}_{k-1}(\Psi_{2},\dots,\Psi_{k}),\Psi_{k+1}),A)))\\
    &=-m_{2}(A,m_{2}(\Psi_{1},\mu^{(1)}_{k}(\Psi_{2},\dots,\Psi_{k+1}))).
\end{align}
Since the relation \eqref{cyc:mu1} has already been established for $k=1$, it follows by induction that \eqref{mucyc} holds for all $k\ge1$.

Using this identity, we now examine the cyclicity condition \eqref{redcyc}. We compute
\begin{align}
    \omega(\mu^{(1)}_{k}(\Psi_{1},\dots,\Psi_{k}),\Psi_{k+1})&=-\omega(m_{2}(A,m_{2}(\Psi_{1},\mu^{(1)}_{k-1}(\Psi_{2},\dots,\Psi_{k}))),\Psi_{k+1})\\
    &=\omega(A,m_{2}(m_{2}(\Psi_{1},\mu^{(1)}_{k-1}(\Psi_{2},\dots,\Psi_{k})),\Psi_{k+1}))\\
    &=-\omega(m_{2}(m_{2}(\Psi_{1},\mu^{(1)}_{k-1}(\Psi_{2},\dots,\Psi_{k})),\Psi_{k+1}),A)\\
    &=-\omega(\Psi_{1},m_{2}(m_{2}(\mu^{(1)}_{k-1}(\Psi_{2},\dots,\Psi_{k}),\Psi_{k+1}),A))\\
    &=-\omega(\Psi_{1},\mu^{(1)}_{k}(\Psi_{2},\dots,\Psi_{k+1})).
\end{align}
This shows that $\mu^{(1)}_{k}$ satisfies the cyclicity condition. In particular, including the case $k=1$, we conclude that $\mu^{(1)}_{k}$ is cyclic for all $k\ge0$, namely,
\begin{equation}
    0=\omega(\mu^{(1)}_{k}(\Psi_{1},\dots,\Psi_{k}),\Psi_{k+1})+\omega(\Psi_{1},\mu^{(1)}_{k}(\Psi_{2},\dots,\Psi_{k+1})).
\end{equation}

Finally, we determine the coderivations $\vb*{\mu}^{(k)}$ for $k\ge2$ so that
\eqref{condhigher} is satisfied. Since we have already constructed the coderivation $\vb*{\mu}^{(1)}$ obeying \eqref{cond}, the condition \eqref{condhigher} can be rewritten as
\begin{equation}
    0=\comm{\vb*{\mu}^{(2)}}{\vb{m}^{\mathrm{tv}}}+\frac{1}{2}\comm{\vb*{\mu}^{(1)}}{\vb{m}_{0}}.
\end{equation}
If the following holds,
\begin{equation}
    0=\comm{\vb*{\mu}^{(1)}}{\vb{m}_{0}},
\end{equation}
we may consistently choose all higher coderivations $\vb*{\mu}^{(k)}$ for $k\ge2$ to vanish. In terms of the multilinear maps, this condition is equivalent to
\begin{equation}
    0=\mu^{(1)}_{n+1}(\hat{m}_{0},\Psi_{1},\dots,\Psi_{n})+(-1)^\abs{\Psi_{1}}\mu^{(1)}_{n+1}(\Psi_{1},\hat{m}_{0},\dots,\Psi_{n})+\dots+(-1)^{\abs{\Psi_{1}}+\dots+\abs{\Psi_{n}}}\mu^{(1)}_{n+1}(\Psi_{1},\dots,\Psi_{n},\hat{m}_{0}).
\end{equation}

We will show that the above condition holds under an assumption
\begin{equation}
    m_{2}(A,A)=0.
\end{equation}
Although it is not clear whether this assumption is valid in general, the homotopy operator for any Okawa-type tachyon vacuum solution \cite{Okawa:2006vm} as well as for the Takahashi-Tanimoto tachyon vacuum solution \cite{Takahashi:2002ez,Inatomi:2011xr} is known to satisfy this condition. For these analytic solutions, this property follows from the structure of the subalgebra used in their construction. In particular, the relevant subalgebras do not contain states with ghost number less than $-1$, which implies $m_{2}(A,A)=0$. Therefore, in what follows, we restrict ourselves to tachyon vacuum solutions whose homotopy operator $A$ obeys $m_{2}(A,A)=0$\footnote{A similar assumption that the homotopy operator squares to zero is also made in \cite{Erler:2019fye}.}. In addition, we impose the further assumption
\begin{equation}
    m_{2}(m_{2}(\hat{m}_{0},\hat{m}_{0}),m_{2}(A,A))=0.\label{assume:m0}
\end{equation}
In general, $m_{2}(\hat{m}_{0},\hat{m}_{0})$ is divergent. However, we require that this divergence is softer, in the sense that its product with $m_{2}(A,A)$ vanishes as above. Then, using
\begin{align}
    \mu^{(1)}_{k+1}(\Psi_{1},\dots,\Psi_{k},\hat{m}_{0})&=m_{2}(m_{2}(\mu^{(1)}_{k}(\Psi_{1},\dots,\Psi_{k}),\hat{m}_{0}),A)\\
    &=-(-1)^{\abs{\Psi_{1}}+\dots+\abs{\Psi_{k}}}m_{2}(m_{2}(\hat{m}_{0},\mu^{(1)}_{k}(\Psi_{1},\dots,\Psi_{k})),A)\\
    &=(-1)^{\abs{\Psi_{1}}+\dots+\abs{\Psi_{k}}}m_{2}(\hat{m}_{0},m_{2}(\mu^{(1)}_{k}(\Psi_{1},\dots,\Psi_{k}),A))\\
    &=(-1)^{\abs{\Psi_{1}}+\dots+\abs{\Psi_{k}}}m_{2}(\hat{m}_{0},m_{2}(m_{2}(m_{2}(\mu^{(1)}_{k-1}(\Psi_{1},\dots,\Psi_{k-1}),\Psi_{k}),A),A))\\
    &=m_{2}(\hat{m}_{0},m_{2}(m_{2}(\mu^{(1)}_{k-1}(\Psi_{1},\dots,\Psi_{k-1}),\Psi_{k}),m_{2}(A,A)))\\
    &=m_{2}(\hat{m}_{0},m_{2}(m_{2}(\dots(m_{2}(\hat{m}_{0},A),\Psi_{1},\dots,\Psi_{k-1}),\Psi_{k}),m_{2}(A,A)))\\
    &=m_{2}(m_{2}(m_{2}(\dots(m_{2}(\hat{m}_{0},\hat{m}_{0}),A),\Psi_{1}),\dots,\Psi_{k}),m_{2}(A,A)),
\end{align}
we immediately obtain, under the assumption \eqref{assume:m0},
\begin{equation}
    \mu_{k+1}(\Psi_{1},\dots,\Psi_{k},\hat{m}_{0})=0.
\end{equation}
Similarly, consider the case where $\hat{m}_{0}$ appears at an intermediate position. By repeatedly applying the recursive definition of $\mu^{(1)}$,
we obtain
\begin{align}
    &\mu^{(1)}_{k+1}(\Psi_{1},\dots,\Psi_{n},\hat{m}_{0},\Psi_{n+1},\dots,\Psi_{k})\\
    &=m_{2}(m_{2}(\mu^{(1)}_{k}(\Psi_{1},\dots,\Psi_{n},\hat{m}_{0},\Psi_{n+1},\dots,\Psi_{k-1}),\Psi_{k}),A)\\
    &=m_{2}(m_{2}(m_{2}(m_{2}(\mu^{(1)}_{k-1}(\Psi_{1},\dots,\Psi_{n},\hat{m}_{0},\Psi_{n+1},\dots,\Psi_{k-2}),\Psi_{k-1}),A),\Psi_{k}),A)\\
    &\vdots\\
    &=m_{2}(m_{2}(\dots(m_{2}(\mu_{n+1}(\Psi_{1},\dots,\Psi_{n},\hat{m}_{0}),\Psi_{n+1}),A),\dots,\Psi_{k}),A)\\
    &=0.
\end{align}
This also implies
\begin{equation}
    \mu^{(1)}_{k+1}(\Psi_{1},\dots,\Psi_{n},\hat{m}_{0},\Psi_{n+1},\dots,\Psi_{k})=0.
\end{equation}
Hence we conclude that
\begin{equation}
    0=\comm{\vb*{\mu}^{(1)}}{\vb{m}_{0}}.
\end{equation}
As a result, all higher coderivations $\vb*{\mu}^{(k)}$ for $k\ge2$ can consistently be taken to vanish.

Therefore, the coderivation
\begin{equation}
    \vb*{\mu}=h\vb*{\mu}^{(1)}
\end{equation}
satisfies all the required conditions. The fields $\Psi(t)$ and $\Psi'(t)$ are related by
\begin{align}
    \Psi(t)&=\pi_{1}e^{\vb*{\mu}}\frac{1}{1-\Psi'(t)}\\
    &=\Psi'(t)+\pi_{1}\vb*{\mu}\frac{1}{1-\Psi'(t)}+\frac{1}{2}\pi_{1}\vb*{\mu}^{2}\frac{1}{1-\Psi'(t)}+\dots.
\end{align}
Under the assumption \eqref{assume:m0}, we obtain
\begin{align}
    &\mu^{(1)}_{k}(\Psi_{1},\dots,\Psi_{l},\mu^{(1)}_{n}(\Psi_{l+1},\dots,\Psi_{l+n}),\Psi_{l+n+1},\dots,\Psi_{k+n-1})\\
    &=m_{2}(m_{2}(\dots(m_{2}(m_{2}(m_{2}(\mu^{(1)}_{l}(\Psi_{1},\dots,\Psi_{l}),A),\mu^{(1)}_{n}(\Psi_{l+1},\dots,\Psi_{l+n})),A),\Psi_{l+n+1}),\dots,\Psi_{l+n-1}),A)\\
    &=m_{2}(m_{2}(\dots(m_{2}(m_{2}(m_{2}(\mu^{(1)}_{l}(\Psi_{1},\dots,\Psi_{l}),A),\\
    &\qquad m_{2}(m_{2}(\mu^{(1)}_{n-1}(\Psi_{l+1},\dots,\Psi_{l+n-1}),\Psi_{l+n}),A)),A),\Psi_{l+n+1}),\dots,\Psi_{l+n-1}),A)\\
    &=m_{2}(m_{2}(\dots(m_{2}(m_{2}(\mu^{(1)}_{l}(\Psi_{1},\dots,\Psi_{l}),A),\\
    &\qquad m_{2}(m_{2}(\mu^{(1)}_{n-1}(\Psi_{l+1},\dots,\Psi_{l+n-1}),\Psi_{l+n}),m_{2}(A,A))),\Psi_{l+n+1}),\dots,\Psi_{l+n-1}),A)\\
    &=0.
\end{align}
Therefore we conclude that
\begin{equation}
    \vb*{\mu}^{2}=0.
\end{equation}
As a consequence, the field redefinition reduces to
\begin{align}
    \Psi(t)&=\Psi'(t)+\pi_{1}\vb*{\mu}\frac{1}{1-\Psi'(t)}\\
    &=\Psi'(t)+h\qty(\mu^{(1)}_{0}+\mu^{(1)}_{1}(\Psi'(t))+\mu^{(1)}_{2}(\Psi'(t),\Psi'(t))+\dots).
\end{align}
Imposing $\Psi'(0)=0$, we have $\Psi(0)=F_{0}$, where
\begin{align}
    F_{0}&=\pi_{1}e^{\vb*{\mu}}\\
    &=h\mu^{(1)}_{0}.
\end{align}
As a result, we find that
\begin{equation}
    \int_{0}^{1}\dd{t}\bra{\omega}\pi_{1}\vb*{\partial}_{t}\frac{1}{1-\Psi(t)}\otimes\pi_{1}\vb{m}^{h,\mathrm{tv}}\frac{1}{1-\Psi(t)}=\int_{0}^{1}\dd{t}\bra{\omega}\pi_{1}\vb*{\partial}_{t}\frac{1}{1-\Psi'(t)}\otimes\pi_{1}\vb{m}^{\mathrm{tv}}\frac{1}{1-\Psi'(t)},
\end{equation}
or equivalently
\begin{equation}
    S^{\mathrm{tv}}_{h}[\Psi]-S^{\mathrm{tv}}_{h}[F_{0}]=S^{\mathrm{tv}}_{0}[\Psi'].
\end{equation}

At first sight, this result appears to suggest that the deformed theory is equivalent to Witten's theory. However, this conclusion cannot be correct. As we will show in the next section, the above field redefinition is in fact not well-defined.

\section{Transfer of classical solution\label{star}}
The result obtained in the previous section is universal in the sense that it relies only on the cyclic $A_{\infty}$ relations together with the assumption \eqref{assume:m0}. Nevertheless, the conclusion appears counterintuitive, and the derivation in the homotopy algebra language is complicated by nontrivial sign factors. In this section, we reexamine the same result using the star product notation, where the structure becomes more transparent.

We adopt the following conventions,
\begin{align}
    \omega(\Psi_{1},\Psi_{2})&=-(-1)^{\mathrm{gh}(\Psi_{1})}\Tr(\Psi_{1}\Psi_{2}),\\
    m^{\mathrm{tv}}_{1}(\Psi)&=Q_{\mathrm{tv}}\Psi,\\
    m_{2}(\Psi_{1},\Psi_{2})&=-(-1)^{\mathrm{gh}(\Psi_{1})}\Psi_{1}\Psi_{2},\\
    m_{0}&=h\mathcal{V},
\end{align}
where $\mathcal{V}$ is an on-shell closed string vertex operator. For notational simplicity, we suppress the explicit star symbol in the product. With these conventions, the action takes the form
\begin{equation}
    S^{\mathrm{tv}}_{h}=\Tr(\frac{1}{2}\Psi Q_{\mathrm{tv}}\Psi+\frac{1}{3}\Psi^{3}+h\Psi\mathcal{V}),
\end{equation}
where we set the open string coupling constant to one. Also, we also express the multilinear maps $\mu^{(1)}_{k}$ in the star product notation. Using the definitions introduced above, we obtain
\begin{align}
    \mu^{(1)}_{0}&=m_{2}(\hat{m}_{0},A)=-\mathcal{V}A,\\
    \mu^{(1)}_{1}(\Psi)&=m_{2}(m_{2}(\mu^{(1)}_{0},\Psi),A)=(-1)^{\mathrm{gh}(\Psi)}m_{2}(\mu^{(1)}_{0},\Psi)A=-(-1)^{\mathrm{gh}(\Psi)}\mathcal{V}A\Psi A,
\end{align}
and for $k\ge0$,
\begin{align}
    \mu^{(1)}_{k+1}(\Psi_{1},\dots,\Psi_{k+1})&=-m_{2}(A,m_{2}(\Psi_{1},\mu^{(1)}_{k}(\Psi_{2},\dots,\Psi_{k+1})))\\
    &=-Am_{2}(\Psi_{1},\mu^{(1)}_{k}(\Psi_{2},\dots,\Psi_{k+1}))\\
    &=(-1)^{\mathrm{gh}(\Psi_{1})}A\Psi_{1}\mu^{(1)}_{k}(\Psi_{2},\dots,\Psi_{k+1})\\
    &=(-1)^{\mathrm{gh}(\Psi_{1})+\mathrm{gh}(\Psi_{2})}A\Psi_{1}A\Psi_{2}\mu^{(1)}_{k-1}(\Psi_{3},\dots,\Psi_{k+1})\\
    &\vdots\\
    &=-(-1)^{\mathrm{gh}(\Psi_{1})+\dots+\mathrm{gh}(\Psi_{k+1})}\mathcal{V}A\Psi_{1}A\dots A\Psi_{k+1}A.
\end{align}
Hence, the relation between $\Psi$ and $\Psi'$ takes the simple form
\begin{align}
    \Psi&=\Psi'-h\mathcal{V}A+h\mathcal{V}A\Psi'A-h\mathcal{V}A\Psi'A\Psi'A+\dots\\
    &=\Psi'-h\mathcal{V}\frac{1}{1+A\Psi'}A,
\end{align}
and the inverse relation is given by
\begin{equation}
    \Psi'=\Psi+h\mathcal{V}\frac{1}{1+A\Psi}A.
\end{equation}
Using this relation, we formally obtain
\begin{equation}
    \Tr(\frac{1}{2}\Psi Q_{\mathrm{tv}}\Psi+\frac{1}{3}\Psi^{3}+h\Psi\mathcal{V})+\frac{1}{2}h^{2}\Tr(A\mathcal{V}^{2})=\Tr(\frac{1}{2}\Psi'Q_{\mathrm{tv}}\Psi'+\frac{1}{3}\Psi'^{3}).
\end{equation}
In order to derive the relation between the actions around the perturbative vacuum, we write
\begin{equation}
    \Psi=\Psi'-h\mathcal{V}\frac{1}{1+A(\Psi'-\Psi_{\mathrm{tv}})}A\qc\Psi'=\Psi+h\mathcal{V}\frac{1}{1+A(\Psi-\Psi_{\mathrm{tv}})}A.
\end{equation}
Accordingly, the actions formally satisfy
\begin{equation}
    \Tr(\frac{1}{2}\Psi Q\Psi+\frac{1}{3}\Psi^{3}+h\Psi\mathcal{V})-h\Tr(\Psi_{\mathrm{tv}}\mathcal{V})+\frac{1}{2}h^{2}\Tr(A\mathcal{V}^{2})=\Tr(\frac{1}{2}\Psi'Q_{\mathrm{tv}}\Psi'+\frac{1}{3}\Psi'^{3}).
\end{equation}
However, we emphasize that these relations keep only at a formal level. The field redefinition is singular because the inverse factors $(1+A\Psi)^{-1}$ and $(1+A\Psi')^{-1}$ are not well-defined in general. Therefore, this field redefinition does not imply that Witten's theory is equivalent to the deformed theory.

Nevertheless, it suggests that classical solutions of Witten's theory can be transferred to classical solutions of the theory deformed by the Ellwood invariant. Suppose that $\Psi_{\mathrm{sol}}$ is a classical solution of Witten's theory,
\begin{equation}
    Q\Psi_{\mathrm{sol}}+\Psi_{\mathrm{sol}}^{2}=0.
\end{equation}
We claim that
\begin{equation}
    \Psi_{h,\mathrm{sol}}\coloneqq\Psi_{\mathrm{sol}}-h\mathcal{V}\frac{1}{1+A(\Psi_{\mathrm{sol}}-\Psi_{\mathrm{tv}})}A
\end{equation}
solves the equation of motion of the deformed theory,
\begin{equation}
    Q\Psi_{h,\mathrm{sol}}+\Psi_{h,\mathrm{sol}}^{2}+h\mathcal{V}=0.
\end{equation}

However, because the above $\Psi_{h,\mathrm{sol}}$ is generally singular, we must verify whether it satisfies the equation of motion in the strong sense. On the other hand, in the special case where we set $\Psi_{\mathrm{sol}}=\Psi_{\mathrm{tv}}$, we obtain
\begin{equation}
    \Psi_{h,\mathrm{tv}}=\Psi_{\mathrm{tv}}-h\mathcal{V}A,
\end{equation}
which is a regular string field. Indeed, we compute
\begin{align}
    &Q(\Psi_{\mathrm{tv}}-h\mathcal{V}A)+\qty(\Psi_{\mathrm{tv}}-h\mathcal{V}A)^{2}+h\mathcal{V}\\
    &=Q\Psi_{\mathrm{tv}}+\Psi_{\mathrm{tv}}^{2}+h(-Q(\mathcal{V}A)-\Psi_{\mathrm{tv}}\mathcal{V}A-\mathcal{V}A\Psi_{\mathrm{tv}}+\mathcal{V})+h(\mathcal{V}A)^{2}.
\end{align}
Using the fact that $\Psi_{\mathrm{tv}}$ is a classical solution, that $\mathcal{V}$ is on-shell, the defining property of the homotopy operator $A$, and the nilpotency of $\mathcal{V}A$, we conclude that $\Psi_{h,\mathrm{tv}}$ satisfies the equation of motion of the deformed theory. However, since the field redefinition itself is not well-defined, it remains unclear whether $\Psi_{h,\mathrm{tv}}$ should be identified with the tachyon vacuum solution in the deformed theory. Nevertheless, we can verify that the cohomology of $Q_{h,\mathrm{tv}}$ is empty by explicitly constructing a homotopy operator where $Q_{h,\mathrm{tv}}$ is defined as
\begin{align}
    Q_{h,\mathrm{tv}}\Psi&\equiv Q\Psi+(\Psi_{\mathrm{tv}}-h\mathcal{V}A)\Psi-(-1)^{\mathrm{gh}(\Psi)}\Psi(\Psi_{\mathrm{tv}}-h\mathcal{V}A)\\
    &=Q_{\mathrm{tv}}\Psi-h\mathcal{V}(A\Psi-(-1)^{\mathrm{gh}(\Psi)}\Psi A).
\end{align}
Acting on the homotopy operator of Witten's theory and using the assumption $A^{2}=0$, we obtain
\begin{align}
    Q_{h,\mathrm{tv}}A&=Q_{\mathrm{tv}}A-2h\mathcal{V}A^{2}\\
    &=1.
\end{align}
This demonstrates that the cohomology of $Q_{h,\mathrm{tv}}$ is empty. Therefore, the tachyon vacuum solution of Witten's theory can be transferred to a tachyon vacuum solution of the deformed theory.

\section{Summary\label{summary}}
We construct a field redefinition relating Witten's open string field theory to its deformation by the Ellwood invariant. However, this field redefinition is not well-defined. Consequently, the resulting relation is purely formal and does not imply an equivalence between Witten's theory and the deformed theory.

Although the field redefinition is singular, this does not necessarily render it meaningless. In Witten's theory, it is well known that solutions can be related by singular gauge transformations \cite{Erler:2012qn}. While this does not imply physical equivalence, it plays an essential role in the construction of nontrivial solutions, such as multiple-brane solutions \cite{Miwa:2017oxy,Hata:2019dwu,Hata:2019ybw}. Likewise, even when the field redefinition is singular, this observation allows us to formally transfer classical solutions of Witten's theory to solutions of the deformed theory.

Of course, we must verify whether the resulting solutions satisfy the equations of motion in the strong sense, and it remains unclear whether their physical interpretation is preserved. Nevertheless, at least in the case of the tachyon vacuum solution, such a transfer is justified since the existence of a homotopy operator can be explicitly confirmed.

In general, a regular solution in Witten's theory is transferred to a singular solution in the deformed theory through our field redefinition, which complicates the analysis. However, the converse possibility is also of interest. A solution that appears singular in Witten's theory may become regular in the deformed theory. If such a case is realized, solutions that have traditionally been considered difficult to handle could prove to be valuable.

\section*{Acknowledgments}
The author would like to thank Nobuyuki Ishibashi for useful comment at early stages of this work. The author also would like to thank Keisuke Konosu for discussions.

\appendix
\section{Homotopy operator\label{homotopyop}}
In this appendix, we introduce the homotopy operator for the tachyon vacuum solution in the language of homotopy algebra.

Let us consider the shifted BRS operator around the tachyon vacuum solution, denoted by $Q_{\mathrm{tv}}$, which carries ghost number one. Suppose that there exists a string field $A$ satisfying
\begin{equation}
    Q_{\mathrm{tv}}(A)=1,
\end{equation}
where 1 is the identity string field with ghost number is zero. If such a string field $A$ exists, any $Q_{\mathrm{tv}}$-closed string field $\Psi$ can be written as
\begin{equation}
    \Psi=1*\Psi=Q_{\mathrm{tv}}(A)*\Psi=Q_{\mathrm{tv}}(A*\Psi),\label{cohotri:gh1}
\end{equation}
or equivalently,
\begin{equation}
    \Psi=\Psi*1=\Psi*Q_{\mathrm{tv}}(A)=(-1)^{\mathrm{gh}(\Psi)}Q_{\mathrm{tv}}(\Psi*A).\label{cohotri:gh2}
\end{equation}
Hence the existence of the homotopy operator $A$ proves that the cohomology of $Q_{\mathrm{tv}}$ vanishes \cite{Ellwood:2006ba}. A string field $A$ satisfying $Q_{\mathrm{tv}}(A)=1$ is referred to as a homotopy operator.

Let us express $m_{2}$ in terms of the star product as
\begin{equation}
    m_{2}(\Psi_{1},\Psi_{2})\coloneqq(-1)^\abs{\Psi_{1}}\Psi_{1}*\Psi_{2}.
\end{equation}
Since the degree is defined as the ghost number minus one, the degrees of the homotopy operator and the identity string field are
\begin{align}
    \abs{A}&=\mathrm{gh}(A)-1=-2,\\
    \abs{1}&=\mathrm{gh}(1)-1=-1.
\end{align}
Using this convention, the equations \eqref{cohotri:gh1} and \eqref{cohotri:gh2} can be rewritten as
\begin{align}
    \Psi&=-m_{2}(1,\Psi)=-m_{2}\qty(m^{\mathrm{tv}}_{1}(A),\Psi)=m^{\mathrm{tv}}_{1}\qty(m_{2}(A,\Psi)),\\
    \Psi&=(-1)^\abs{\Psi}m_{2}(\Psi,1)=(-1)^\abs{\Psi}m_{2}\qty(\Psi,m^{\mathrm{tv}}_{1}(A))=-m^{\mathrm{tv}}_{1}\qty(m_{2}(\Psi,A)).
\end{align}
In particular, the first equalities in the first and second lines reproduce \eqref{def:identitystring}.

\printbibliography
\end{document}